 \documentclass[manuscript]{aastex}

\def\lesssim{\mathrel{\hbox{\rlap{\hbox{\lower4pt\hbox{$\sim$}}}\hbox{$<$}}}}
\def\gtrsim{\mathrel{\hbox{\rlap{\hbox{\lower4pt\hbox{$\sim$}}}\hbox{$>$}}}}
\def\kf{\left(\frac{k/f}{0.02}\right)}

\usepackage{graphicx}
\usepackage{color}
\usepackage{natbib}
\usepackage{amsmath}
\begin{document}

\title{Jets from Tidal Disruptions of Stars by Black Holes}
\shorttitle{Jets from Tidal Disruptions}

\author{Julian H. Krolik}
\affil{Physics and Astronomy Department\\
Johns Hopkins University\\ 
Baltimore, MD 21218}

\and

\author{Tsvi Piran}
\affil{Racah Institute of Physics\\
The Hebrew University of Jerusalem\\ 
Jerusalem 91904, Israel}

\email{jhk@jhu.edu; tsvi@phys.huji.ac.il}

\begin{abstract}

Tidal disruption of main sequence stars by black holes has generally been thought
to lead to a signal dominated by UV emission.  If, however, the black hole spins
rapidly and the poloidal magnetic field intensity on the black hole horizon is
comparable to the inner accretion disk pressure, a powerful jet may form whose luminosity 
can easily exceed the thermal UV luminosity.  When the jet beam points at Earth, its
non-thermal luminosity can dominate the emitted spectrum.  The thermal and non-thermal
components decay differently with time.  In particular, the thermal emission should
remain roughly constant for a significant time after the period of maximum
accretion, beginning to diminish only after a delay, whereas after the peak accretion
rate, the non-thermal jet emission decays, but then reaches a plateau.  Both transitions
are tied to a characteristic timescale $t_{\rm Edd}$ at which the accretion rate
falls below Eddington.  Making use of this timescale in a new parameter-inference
formalism for tidal disruption events with significant emission from a jet, we analyze
the recent flare source Swift~J2058.  It is consistent with an event in which a main
sequence solar-type star
is disrupted by a black hole of mass $\sim 4 \times 10^7 M_{\odot}$.  The beginning
of the flat phase in the non-thermal emission from this source can possibly be seen
in the late-time lightcurve.  Optical photometry over the first $\simeq 40$~d of this flare
is also consistent with this picture, but is only weakly constraining because the
bolometric correction is very uncertain.  We suggest that future searches for main
sequence tidal disruptions use methods sensitive to jet radiation as well as to thermal
UV radiation.

\end{abstract}


\section{Introduction}

Now that we know that supermassive black holes frequently can be found in the
centers of galaxies \citep{kormendyrichstone95,gultekin09}, we can also expect
that ordinary stars occasionally pass near enough to them to be tidally disrupted.
Theoretical estimates of the rate at which this might happen range from
$3 \times 10^{-5} M_{BH,6}^{0.45}$~galaxy$^{-1}$~yr$^{-1}$ \citep{brockamp11} to
$6.5 \times 10^{-4} M_{BH,6}^{-0.25}$~galaxy$^{-1}$~yr$^{-1}$ \citep{wangmerritt04}.
Most observational estimates are in the same ballpark \citep{gezari08,esquej08,maksym10},
although \cite{donley02} estimated a rate of X-ray flares as low as
$\sim 10^{-5}$~galaxy$^{-1}$~yr$^{-1}$.

More than twenty years ago, \cite{rees88} outlined the basic dynamical processes
relevant to these events.  The most likely trajectory for a tidal disruption is
one in which the star approaches on a nearly parabolic orbit.  
When the star is disrupted, its different parts acquire binding energies
that span a very large range.  The material of the star begins returning
to the vicinity of the black hole after a delay corresponding to the orbital period
of the most tightly bound matter.  If the disrupted star has a uniform
distribution in orbital binding energy per unit mass across that large range,
matter continues to fall in at a rate $dM/dt \propto t^{-5/3}$ \citep{phinney89}.
This matter is captured into an accretion disk. Within this disk, gravitational
energy is dissipated and the heat is radiated in the usual quasi-thermal
fashion.  At the peak accretion rate, the luminosity would be super-Eddington \citep{ulmer99};
if it is thermally radiated by a disk of area corresponding
to a few gravitational radii of the black hole, the associated temperature would
be in the EUV or perhaps the soft X-ray band. Consequently, most searches for 
such events hitherto have been carried out in the EUV region.

Many of these expectations were upended by the remarkable object Swift J1644+57
\citep{levan11,burrows11}.   During its time of peak luminosity (the first $\simeq 2$~d
of the outburst), its lightcurve, with its sequence of flares lasting $\sim 1000$~s
alternating with quiescent periods $\sim 5 \times 10^4$~s long, bore little resemblance
to the prediction of a rise-time of $\sim 10^6$~s followed by a smooth roll-over
to a $t^{-5/3}$ fall-off.  Moreover, whereas all expectations were for the emission
to peak in the EUV, the greatest part of the luminosity was
carried in photons with energies $\sim 100$~keV, indicating electron energies far greater
than the expected $\sim 100$~eV, while the shape of that spectrum ($\nu F_\nu$
gradually rising from 2~keV to 100~keV) certainly did not suggest any variety
of thermal emission.  In fact, the most successful emission models invoke
non-thermal synchro-Compton radiation from a relativistic jet \citep{bloom11,burrows11}.

If most of the radiation actually comes from a relativistic jet, then clearly some
of the conventional assumptions must be revised.  The observed radiation has nothing
to do with quiescent radiative cooling of an optically thick accretion flow.  Indeed,
if the jet is powered by something resembling the Blandford-Znajek mechanism,
its real source of energy is rotation of the black hole; the only function of the
mass accretion is to support the magnetic field threading the black hole's horizon.

We have recently shown \citep{kp11} that a possible solution to all these
conundrums is that this event was caused by tidal
disruption not of a main sequence star, but of a white dwarf, and
the disruption takes place over several passes, not in a single go.  Most importantly
for the question examined in that paper, a white dwarf disruption results in
depositing so much mass in such a small region that the accretion flow is
immensely optically thick.  As a result, the matter's cooling time is far longer
than its inflow time.  In such a state of affairs, thermal radiation from the
accretion flow is suppressed, and the disk maintains a high pressure, one
capable of confining a strong magnetic field close to the black hole. 

Our goal in this paper is to inquire whether a similar state of affairs could
occur when the victim of tidal disruption is a main sequence star.  We estimate
here both the thermal emission of the disk and the jet power. The first is estimated
using standard accretion disk theory.
%and as the Eddington luminosity in the super Eddington regime.
The latter is estimated in terms of the Blandford-Znajek mechanism, using the disk
pressure near the ISCO as a measure of the black hole's magnetic field.
%(see also Moderski and Sikora, 1996??): MS96 simply rephrases the formula
%based on a small a/M expansion in BZ77. 
Although a tidal disruption of a main sequence star takes place on a much larger
lengthscale than the disruption of a white dwarf, for most of the relevant parameter
space a similar situation holds: the accretion is super-Eddington, thermal radiation
is suppressed, and conditions for emergence of a strong jet are established.

Our work is complementary to two related efforts.  \cite{giannios11} investigated possible
radio emission from a jet receiving a fixed (small) fraction of the accretion energy released
by accreting tidally disrupted matter. \cite{leizhang11} have suggested a similar picture,
but approach it rather differently.  In particular, they use thin disk approximations for both
the sub- and super-Eddington regimes, their scaling with black hole mass does not include
the relation between disk thickness and accretion rate in the radiation-dominated sub-Eddington
phase, and they do not discuss the luminosity of the thermal disk.  We begin in \S~\ref{sec:tidal}
with a brief discussion of tidal disruption physics.  We then discuss accretion dynamics
in this context and the jet and disk outputs in \S~\ref{sec:jet_disk}.  In \S~\ref{sec:j2058},
we show how this approach can be used to constrain a number of otherwise-unknown parameters of
tidal disruptions and apply this method to Swift~J2058 \citep{cenko11}, a second example of a
jet-dominated tidal disruption.  We summarize our results in \S~\ref{sec:conc}.

\section{Tidal Disruption} 
\label{sec:tidal}

In order for the tidal gravity of a black hole to strongly affect a star,
the pericenter of the star's orbit cannot be much larger than
\begin{equation}
R_T \simeq 50 (k/f)^{1/6} {\cal M}_*^{2/3 -\xi} M_{BH,6}^{-2/3} R_g,
\end{equation}
where $k$ is the apsidal motion constant (determined by the star's radial density
profile) and $f$ is its binding energy in units of $GM_*^2/R_*$ \citep{phinney89}.
Here ${\cal M}_*$ is the mass of the star in solar units, $M_{BH,6}$
is the black hole's mass in units of $10^6 M_\odot$, and $R_g$ is the black
hole's gravitational lengthscale $GM_{BH}/c^2$.  We have approximated the main
sequence mass-radius relation by $R_* \approx R_\odot M_*^{(1-\xi)}$;
$\xi \simeq 0.2$ for $0.1 < {\cal M}_* \le 1$, but increases to $\simeq 0.4$ for
$1 < {\cal M}_* < 10$ \citep{kipp94}.
The analogous expression for non-main sequence stars can be found by substituting
the appropriate mass-radius relation; for example, for white dwarfs with
masses well below the Chandrasekhar mass,
$\xi=4/3$, and the radius of a $1 M_{\odot}$ star is $\simeq 0.011 R_{\odot}$.
The ratio $k/f \simeq 0.02$ for radiative stars, but is $\simeq 0.3$ for convective stars
\citep{phinney89}.  Although the ratio $k/f$ appears in the expression for $R_T$ raised
to only the $1/6$ power, and would therefore appear to be an innocuous correction factor
of order unity, it can be quantitatively significant because the timescale
for the flare is $\propto R_T^3$  and the contrast between the values of $k/f$ for
radiative and convective stars is $\simeq 15$.  In numerical estimates, we scale $k/f$
to the value for fully radiative stars because this is a reasonable approximation
for main sequence stars with $0.4 M_{\odot} < M < 10 M_{\odot}$ \citep{kipp94}.

The work done by the tides stretching and rotating the star is removed from the orbit,
so on average, the star's material becomes bound to the black hole with a specific
binding energy of order the star's original specific binding energy $\sim GM_*/R_*$.
Because the semi-major axis $a$ of an orbit with binding energy $E_B$ is $GM_{BH}/(2E_B)$
when $M_* \ll M_{BH}$, the semi-major axis of material with the mean energy is
much larger than $R_T$,
\begin{equation}
a_{\rm mean} \sim (1/2) (k/f)^{-1/6}(M_{BH}/M_*)^{2/3} R_T.
\end{equation}
However, when the star is disrupted, its different parts acquire binding energies
that span a very large range.  Suppose that the actual pericenter is $R_p = \beta R_T$,
where $\beta \lesssim 1$ is called the ``penetration factor".
%(unfortunately, there is no standard notation for this quantity; for example, \cite{lkp09}
%define it this way, while \cite{lodatorossi11} define it as this quantity's inverse).
The most bound matter may have a binding energy as great as $\sim GM_{BH}R_*/R_p^2$,
leading to a semi-major axis only
\begin{equation}\label{eq:porb_min}
a_{\rm min} \sim (1/2)\beta^{2} (k/f)^{1/6}M_{BH,6}^{2/3}{\cal M}_*^{1/3 -\xi} R_T,
\end{equation}
much smaller than $a_{\rm mean}$.  Tidal torques during the initial stage of the encounter
could also lead to a somewhat larger binding energy for the most bound material \citep{rees88};
in terms of the discussion in this paper, the effects are indistinguishable from those
due to a decrease in $\beta$.

On the assumption that the disrupted star has a uniform distribution in orbital binding
energy per unit mass, matter returns to the region $\sim R_p$ from the black hole
at a rate $dM/dt \propto (t/t_0)^{-5/3}$ \citep{phinney89}.  The characteristic timescale
$t_0$ for initiation of this power-law accretion rate is the orbital period for the
most bound matter
\begin{equation}\label{eqn:porbamin}
P_{\rm orb}(a_{\rm min}) \simeq 5 \times 10^5 {\cal M}_*^{(1-3\xi)/2}
   M_{BH,6}^{1/2}\kf^{1/2} \beta^{3}\hbox{~s},
\end{equation}
the period for an orbit with semi-major axis $a_{\rm min}$.  More detailed calculations
(e.g., \cite{lkp09}) indicate that, almost independent of the star's internal structure,
the mass return rate does follow this behavior at late times, but at early times
the rate rises rapidly after a delay $\sim P_{\rm orb}(a_{\rm min})$ and then gradually
rolls over to the $t^{-5/3}$ proportionality, transitioning more slowly when the
star's pre-disruption structure is more centrally concentrated.

In this picture, once the matter returns to the vicinity of $R_p$ it is captured
into an accretion disk whose inflow time $t_{\rm in} \ll P_{\rm orb}(a_{\rm min})$.
As the matter moves inward through this disk, there is local dissipation of the
conventional accretion disk variety, and the heat is radiated in the usual quasi-thermal
fashion.  At the peak accretion rate estimated by \cite{lkp09} the luminosity would
be
\begin{equation}\label{eq:charlum}
L_{\rm peak}\sim 1 \times 10^{47} (\eta/0.1) {\cal M}_*^{(1+3\xi)/2}
    M_{BH,6}^{-1/2}\kf^{-1/2}\beta^{-3}\hbox{~erg/s}
\end{equation}
for radiative efficiency $\eta$.  In Eddington units, this luminosity is
$\simeq 800\beta^{-3}(\eta/0.1){\cal M}_*^{(1+3\xi)/2}M_{BH,6}^{-3/2}$
for radiative stars and $\simeq 4$ times smaller for convective stars (see also
\cite{ulmer99,strubbe09}); as we will discuss in \S~\ref{sec:jet_disk}, super-Eddington
accretion may result in only a fraction of this inflow rate reaching the black hole.
If such a luminosity were thermally radiated by a disk of area $\sim O(10)\pi R_g^2$,
the associated temperature would be
\begin{equation}
T_{\rm char} \sim 6\times 10^6 {\cal M}_*^{(1+3\xi)/8} M_{BH,6}^{-5/8}
      \kf^{-1/8}\beta^{-3/4}\hbox{~K},
\end{equation}
implying that most of the radiation would emerge in the soft X-ray band.

\section{Accretion dynamics and the jet and disk powers}
\label{sec:jet_disk}
We begin by estimating conditions at the time of peak accretion rate.  \cite{lkp09}
estimate that $\sim 1/3$ of the star's mass should arrive at $R_p$
after a delay $\sim P_{\rm orb}(a_{\rm min})$.  Because $\dot M$ falls $\propto t^{-5/3}$
thereafter, $dM/d\ln t$ is greatest during this period, so this is both the period
of greatest luminosity and the period in which
most of the energy of the entire event is radiated.  As the matter streams in, its
orbital velocity is roughly the free-fall velocity at $R_p$.  Collision with
any mass in a more circular orbit involves velocity differences of order the
orbital speed, so the post-shock temperature is comparable to the virial temperature.
Following conventional accretion disk theory in the supposition that the inflow rate is
controlled by angular momentum transport and writing the vertically-integrated stress as
some number $\alpha$ times the similarly vertically-integrated pressure, we estimate the
inflow time at $R_p$ as
\begin{mathletters}
\begin{eqnarray}
t_{\rm in} \sim \alpha^{-1} (R_p/H)^2 (R_T/R_G)^{3/2} (R_g/c) \kf^{1/4}\beta^{3/2}\\
           \sim 2 \times 10^4 (\alpha/0.1)^{-1}(k/f)^{1/4} (R_p/H)^2 
           {\cal M}_*^{1-(3/2)\xi}\beta^{3/2}\hbox{~s}.
\end{eqnarray}
\end{mathletters}
The quantity $H$ is the disk thickness, which cooling may
reduce to be less than $R_p$, while $\alpha$ is the usual ratio of stress to pressure.
Note that $t_{\rm in}$ depends only on $M_*$, and not at all on
$M_{BH}$, because it is fundamentally a dynamical time, and at the tidal
radius that is determined by the dynamical time of the star $\sim (R_*^3/GM_*)^{1/2}$.
Because $t_{\rm in} \ll P_{\rm orb}(a_{\rm min}$, the mass held at $R_p$ is determined by a
balance between the slowly-changing rate at which returning tidal streams
deliver mass and the more rapid inflow rate.  The characteristic Thomson
optical depth from the middle of the material to the outside at $R_p$ is then
\begin{equation}
\tau_T \sim 3 \times 10^4 (\alpha/0.1)^{-1} (R_p/H)^2 {\cal M}_*^{(1+12\xi)/6}
           M_{BH,6}^{-7/6} \kf^{-1/12}\beta^{-7/2},
\end{equation}
and the cooling time is
\begin{equation}
t_{\rm cool} \sim \tau_T H/c \sim 4 \times 10^6 (\alpha/0.1)^{-1} (R_p/H)
        {\cal M}_*^{(5+6 \xi)/6} M_{BH,6}^{-5/6}\kf^{1/12}\beta^{-5/2}\hbox{~s}.
\end{equation}

The effectiveness of cooling can be measured by comparing $t_{\rm cool}$ to the
inflow time $t_{\rm in}$ at that radius:
\begin{equation}\label{eq:tcool_tin}
t_{\rm cool}/t_{\rm in} \simeq 500 (H/R_p) {\cal M}_*^{-1/6+(5/2)\xi} M_{BH,6}^{-5/6}
          \kf^{-1/6}\beta^{-4}.
\end{equation}
If the initial temperature of the matter arriving in the disk is close to the virial
temperature, it cools efficiently only if the black hole mass is $\sim 10^9 M_{\odot}$
or the star's mass is $\sim 0.1 M_{\odot}$.  Although it is possible for stars on
specific trajectories to be tidally disrupted by spinning black holes with masses
up to $\sim 10^9 M_{\odot}$, the probability of tidal disruption (as opposed to direct
capture) declines rapidly when $M_{BH,6} \sim 100 {\cal M}_*^{1 - (3/2)\xi}$ or more
\citep{kes11}.  For this reason, unless the matter arrives at $R_p$ substantially cooler,
we expect that at the time of peak accretion rate, the inflow will never be able to cool
efficiently.

Another way of viewing this result is to note that photon trapping is expected whenever
the accretion rate is super-Eddington \citep{begelman79,abramowicz88}.  As we have
already seen, the characteristic peak luminosity in these events exceeds the Eddington
luminosity by a factor $\sim 800 (\eta/0.1){\cal M}_*^{(1+3\xi)/2} M_{BH,6}^{-3/2}\kf^{-1/2}\beta^{-3}$.
Thus, to order-of-magnitude accuracy, the dividing line between sub- and
super-Eddington accretion in main sequence tidal disruption events falls at roughly
the same place as the dividing line between the radiatively efficient and photon-trapping
regimes, and both are near the maximum black hole mass at which there is a significant
probability of tidal disruption \citep{ulmer99,strubbe09}.

\subsection{Jet power}

Rotating black holes whose horizons are threaded by large-scale poloidal field
can drive relativistic jets \citep{bz77,mckg04,hk06}.  At the order of magnitude
level, the Poynting luminosity of such a jet can be estimated through a simple
dimensional argument: it must be $\sim c B^2 R_g^2$, where $B^2$ is the
poloidal field intensity on the horizon.  The actual magnitude of the luminosity
can then be written as $L_{\rm jet} = f(a/M) c(B^2/8\pi) R_g^2$, where $f(a/M)$ is
a dimensionless function that should increase with $|a/M|$.  To estimate $B^2$, we follow
\cite{beckwith09}, who demonstrated that the magnetic pressure near the horizon is
generally bounded above by the midplane total pressure near the ISCO and bounded
below by the magnetic pressure at that location.  Our first task is therefore
to estimate the disk pressure near the ISCO.

When the black hole mass is very large (i.e., roughly the maximum black hole mass
capable of disrupting the star before it enters the black hole), the inflow can
be both sub-Eddington and cool efficiently even at the time of peak accretion rate.
If $t_{\rm cool}/t_{\rm in} < 1$ at $R_p$
(eqn.~\ref{eq:tcool_tin}), the matter can cool to the temperature at $R_p$
associated with steady accretion at the prevailing rate.  At smaller radii,
the temperature profile of the accretion flow, as well as its other properties,
can be described by standard stationary disk solutions (e.g., as in \cite{k99}).
Although the accretion rate is sub-Eddington, it is nonetheless high enough that
the flow will certainly be radiation-dominated.  The pressure in the midplane of
the inner disk is then
\begin{equation}
p_{\rm mid} \simeq \frac{2c^2}{\alpha \kappa R_g} (R/R_g)^{-3/2} X(R/R_g),
\end{equation}
where $\kappa$ and $\kappa_T$ are the actual and the Thomson, opacity, respectively, and
the function $X(R/R_g)$ is a place-holder to account for how these scalings are altered
as the ISCO is approached.  Note that $p_{\rm mid}$ in the sub-Eddington
radiation-dominated regime is inversely proportional to $M_{BH}$.  Although the outgoing
thermal radiation flux is proportional to the accretion rate, the optical depth in this
regime is inversely proportional; as a result, the midplane radiation pressure $p_{\rm mid}$
is {\it independent} of the accretion rate \citep{moderski96}.

Conversely, in the low mass limit (which applies to the majority of relevant
black holes), the inflow is both super-Eddington and radiatively
inefficient because photon-trapping prevents effective radiative cooling. 
%In this regime, the theory is less clear.  The ``slim disk" model \citep{abramowicz88} is
%often used to analyze this case, but it depends strongly on the $\alpha$-model, which
%becomes especially dubious near the ISCO \citep{krolik99,g99,khh06,nkh10}.
In this regime, the effects of strong radiation forces and possible radiation-driven
outflows make the theory less clear (consider, e.g., the range of views presented
in \cite{abramowicz88,king03,strubbe09,dotan11,begelman11}).  For our purposes, it is
sufficient to estimate the pressure using dimensional analysis and simple scaling arguments.
Precisely because the gas cannot cool itself, the ratio of total (i.e., radiation plus gas)
pressure to density remains at a level close to the virial temperature as the flow moves
inward, so that, unlike the sub-Eddington radiation pressure-dominated regime, the pressure
when the flow is super-Eddington is proportional to the accretion rate and the flow
configuration is nearly round.  The midplane pressure (again, dominated by
radiation) is then
\begin{equation}
p_{\rm mid} \sim \frac{q\dot m c^2}{\kappa_T R_g} \frac{c_s^2}{\alpha v_{\rm orb}^2}
                 (R/R_g)^{-5/2},
\end{equation}
where $c_s$ is the effective sound speed (including radiation pressure) and
$v_{\rm orb}$ is the speed of a circular orbit at that radius.  Here we have
substituted $\alpha$ for the ratio of inflow speed $v_r$ to orbital speed because
$v_r/v_{\rm orb} \sim \alpha$ when the disk is geometrically thick.
We define $\dot m$ to be the accretion rate delivered to the outer edge of the disk
normalized to Eddington units with {\it unit} efficiency, i.e., $\dot m = \dot M c^2/L_E$
because the uncertain effect of photon-trapping makes the radiative efficiency difficult
to estimate.  When the distinction is important, we will use $\dot m_0$ for the value
of this normalized accretion rate at its peak value.  The parameter $q$ gives the fraction of
$\dot m$ arriving at the black hole; when $\dot m > 1$, it is possible for $q \leq 1$,
but when $\dot m < 1$, in general we expect $q=1$.  Modulo corrections of order unity,
when electron scattering dominates the opacity
and the relevant radii are $\sim R_g$, the only difference between this expression and
the previous one is that this one is also $\propto q\dot m$.

Because our estimates for the total pressure in both the sub- and super-Eddington regimes
are so similar, it is convenient to write an estimate for the jet luminosity in
a single form:
\begin{equation}\label{eq:jetpower}
L_{jet} = \frac{c^3 R_g}{\alpha\beta_h \kappa_T} f(a/M)\begin{cases}
          1 & \dot m \ll 1\\
          q\dot m & q\dot m \gg 1,
\end{cases}
\end{equation}
where $\alpha$ should be interpreted as the usual stress parameter in the sub-Eddington
regime, but only as $v_r/v_{\rm orb}$ in the super-Eddington case.  When
$M_{BH,6}f(a/M)/(\alpha\beta_h)$ = 1 and (in the super-Eddington regime $q\dot m = 1$),
the magnitude of the jet power is $1 \times 10^{43}$~erg~s$^{-1}$.  The quantity
$\beta_h$ is the ratio of the midplane total pressure near the ISCO to the
magnetic pressure in the black hole's stretched horizon.  In very rough terms, we might
expect $\beta_h \alpha \sim 0.1$--1.  Thus, for fixed black hole spin, the power in
the jet is a constant fraction of the Eddington luminosity, independent of accretion
rate, in the sub-Eddington regime, but rises in proportion to the accretion rate
in the super-Eddington regime.  The reason is that at lower accretion rates, the
radiative losses curb any increase in pressure with increasing accretion rate, whereas
once the accretion rate becomes super-Eddington, the photons are trapped in the accretion
flow.

However, there are several subtleties hidden in this estimate.  The first is that
$f(a/M)$ is not well-determined.  \cite{bz77} worked out an expansion in small $a/M$ for
time-steady split-monopolar field distributions.  \cite{tchekh10} developed a
similar expansion in terms of the black hole rotation rate.  Both required
the field rotation rate to be exactly half the black hole rotation
rate.  \cite{tchekh10} found that their expression compared well to axisymmetric
force-free MHD simulations with an imposed magnetic flux function.  Unfortunately,
neither axisymmetry nor an imposed flux function may be a good approximation to
the behavior of time-dependent 3-d relativistic jets whose field configuration is
created self-consistently by accretion dynamics, which, in
turn, may be influenced by black hole spin \citep{hk06}.  It
is therefore difficult to be confident about much more than that jet power should
increase with faster black hole rotation.

The second subtlety is that the relevant field intensity on the horizon is the
time-average of the poloidal component.  When the sign of the vertical field through
the equatorial plane changes frequently, this intensity can be strongly suppressed
\citep{beckwith08}.  Tidal disruptions, however, may be a particularly favorable case
for jet-launching because the extreme elongation of tidal streams might create very
large-scale structures in the field with exactly the sort of coherent direction required
to support powerful and long-lived jets.

Third, up to this point we have been careful to discuss only the two extreme limits:
high-mass, sub-Eddington, radiatively efficient; and low-mass, super-Eddington, and
radiatively inefficient.  That is because the transition between them is not well
understood.  For estimating both the jet luminosity and (later) the emergent thermal
luminosity, the key parameter is the ratio $t_{\rm cool}/t_{\rm in}$.  Several problems
must be solved before its transition between the two extreme limits can be defined.
There may be parameter regimes in which this ratio is $>1$ at $R_p$ if the gas is
deposited there with $T \sim T_{\rm vir}$ even though the temperature in a steady-state
flow at the specified accretion rate is sufficiently cooler that the gas could cool
in less than an inflow time.   On the other hand, there are accretion rates for
which the flow in steady-state is radiatively efficient at $R_p$, but not at smaller
radii.  The quantitative character of the transition is also sensitive to exactly
how the dissipation profile varies with radius near the ISCO.  In view of these
uncertainties, we will adopt here the simplest solution, understanding that it
is only provisional: we will define the transition point by the intersection
between the two expressions for $L_{\rm jet}$; we will adopt a similar prescription
for the thermal luminosity.

Fourth, there is the question of what fraction of the returning mass actually reaches
the vicinity of the black hole, i.e., the correct value of $q$.  This is important in the
super-Eddington regime where $L_{\rm jet} \propto q\dot m$.  If a significant
fraction of $\dot m$ is ejected as a wind (as could well happen during the super-Eddington
period), $q$ could be rather small, and the jet power would be reduced accordingly.

Lastly, jets in Galactic black hole binaries turn off when
$\dot m$ rises toward $\sim 1$ \citep{remmc06}.  On the other hand, the existence of
radio-loud AGN with strong optical/UV continua and emission lines suggests that
larger black holes can somehow support jets even when the accretion rate is near
Eddington.  Luminous quasars are thought to accrete at rates such that
$0.01 \lesssim L/L_E \lesssim 1$ (the lower end of this range favored by, e.g.,
\cite{kelly10}, who estimate black hole masses from the quasar luminosity and
broad-line width; the higher end favored by, e.g., \cite{liu09}, whose black hole
masses come from the bulge dispersion in obscured quasars).  Although these statistics
are dominated by the radio-quiet variety, they likely apply to radio-loud quasars as
well because there are only slight differences between the optical/UV continua of
radio-loud and radio-quiet quasars \citep{richards11}, a fact suggesting that the
inner disks of these two categories of quasar are very similar.  A more direct
indication comes from Narrow Line Seyfert~1 galaxies (NLS1s).  These objects
are thought to accrete at near- or possibly super-Eddington rates
\citep{boller96,boroson02,wang03}.  Recently, it has been found that a fraction of them
are radio-loud \citep{komossa06}, and several of these radio-loud cases are strong
$\gamma$-ray sources \citep{abdo09}; such objects must certainly have strong jets.
Because the black holes in galactic nuclei responsible for tidally disrupting stars
have masses closer to those in AGN than those in Galactic black hole binaries, we will
assume that the AGN example is the guiding precedent here.

In spite of all those caveats, we believe that the fundamental characteristics of this
system make production of a significant relativistic jet very likely when the black
hole spins rapidly.  Moreover, if there is any jet associated directly with the black hole,
on dimensional grounds its luminosity must scale with $c p_{\rm mid} r_g^2$, although
the dimensionless factor multiplying this quantity could be far from unity.

The previous estimate for jet power (eqn.~\ref{eq:jetpower}) posed the issue in
terms of the accretion rate relative to Eddington.  However, this accretion rate
is predicted (approximately) by the dynamics of tidal disruption:
\begin{equation}\label{eq:ljet}
L_{\rm jet} \sim 1 \times 10^{43} \frac{f(a/M)}{\beta_h \alpha}\hbox{~erg/s}
\begin{cases}
8 \times 10^3 q (\dot m/\dot m_0){\cal M}_*^{(1+3\xi)/2}M_{BH,6}^{-1/2}\kf^{-1/2}\beta^{-3}
                       &  M_{BH} \lesssim M_{BH{\rm jet}}\\
M_{BH,6}               &  M_{BH} \gtrsim M_{BH{\rm jet}},
\end{cases}
\end{equation}
where
\begin{equation}
M_{BH\rm jet} = 4 \times 10^8 (\dot m/\dot m_0)^{2/3}{\cal M}_*^{(1+\xi)/3}\kf^{-1/3}
\beta^{-2} M_{\odot}
\end{equation}
In other words, the jet luminosity has a {\it minimum} as a function of $M_{BH}$.
When the black hole mass is smaller, $L_{\rm jet} \propto M_{BH}^{-1/2}$ because
that is the photon-trapping regime, making $p \propto q\dot m$, and tidal disruption
mechanics make $\dot m_0 \propto M_{BH}^{-3/2}$; when the black hole mass is larger,
the accretion flow is in the conventional sub-Eddington regime, and $L_{\rm jet}
\propto M_{BH}$ because $p$ is independent of $\dot M$ but $\propto M_{BH}^{-1}$, while
the black hole area $\propto M_{BH}^2$.  The transition mass between the two regimes, 
$M_{BH\rm jet}$, decreases with decreasing accretion rate. Thus, a system that is
initially in the photon-trapping regime will eventually become sub-Eddington later.
On the other hand, a large black hole system that is sub-Eddington initially will remain
so throughout the event.
Even if only a small fraction of the fallback accretion rate reaches the black
hole (i.e., $q \ll 1$), the jet luminosity would still increase toward smaller black
hole masses provided $q$ scales less rapidly with black hole mass than $\propto M_{BH,6}^{1/2}$.

It is important to recall that at the time of peak accretion,
the black hole mass dividing the low- and high-mass regimes is almost as large
as the very largest black hole mass permitting tidal disruption of main sequence
stars \citep{kes11}.  This conclusion would be weakened somewhat for stars with large
internal convection zones (stars at either the low or high end of the mass
distribution), but $M_{BH {\rm jet}}$ decreases by only a factor $\simeq 2.5$.
At the peak of the flare, therefore, we expect essentially
all events to be effectively in the low-mass, super-Eddington regime.  Only at
later times, as the accretion rate falls, does the high-mass regime become relevant. 

Note also that the jet luminosity is {\it not} directly limited by the Eddington luminosity
because its energy is not directly related to the accretion flow.  The pressure in the
accretion flow confines the magnetic field, but the magnetic field taps the rotational
kinetic energy (i.e., the difference between the total mass and the irreducible mass)
of the black hole.  For the same reason, the jet efficiency measured in rest-mass units,
which is $f(a/M)/(4\pi\beta_h\alpha)$, is likely to be $\sim O(0.1)$, but can in principle
be greater than unity.  

It is also important to emphasize that the jet power does not translate
directly into the photon luminosity we see.  On the one hand, the photon luminosity of
the jet is generally a small fraction $\eta_{\rm jet}$ of the initial Poynting 
luminosity.  In the conditions relevant to blazar jets, for example,
$\eta_{\rm jet} \sim 10^{-2}$--$10^{-1}$ \citep{celottighisellini08}.
On the other hand, whatever photon luminosity emerges will in general be strongly beamed
due to jet collimation and relativistic kinematics:
${\cal B} \sim \min(4\pi/\Delta\Omega,2\Gamma^2)$, where $\Delta \Omega$ is the solid angle
occupied by the jet and $\Gamma$ is its Lorentz factor.  The beaming factor ${\cal B}$ can
easily be $\sim 10^2$.

\subsection{Thermal luminosity}

The thermal photon luminosity behaves differently as a function of black
hole mass.  It can be written as $L_{\rm therm} = \eta q\dot m L_E$, where
we expect that $\eta$ depends on spin for $\dot m < 1$, but declines
once $\dot m$ is large enough that the portion of the accretion
flow where the majority of the luminosity would otherwise emerge enters the
photon-trapping regime.  In this limit, the photon luminosity is capped at
$\sim L_E$.  As we have already discussed in the context of the
jet power, the critical $\dot m$ above which this occurs
is not well-determined.  Thus, we estimate
\begin{equation}\label{eq:ltherm}
L_{\rm therm} \simeq 1.5 \times 10^{44}\hbox{~erg/s}
%\begin{cases}
\left\{\begin{array}{ll}
M_{BH,6}            & M_{BH} \lesssim  M_{BH{\rm therm}}\\
800 (\eta/0.1)(\dot m/\dot m_0){\cal M}_*^{(1+3\xi)/2} M_{BH,6}^{-1/2}\kf^{-1/2}\beta^{-3}  & 
               M_{BH} \gtrsim  M_{BH{\rm therm}}.
\end{array}\right.
%\end{cases}
\end{equation}
%As in the case of $L_{\rm jet}$, our numerical estimate for the super-Eddington
%case assumes that most of the available mass inflow does reach the black hole.  It
%is of less importance here because the associated photon-trapping likely limits
%the luminosity to roughly the Eddington level even if all the bound matter from
%the tidal disruption reaches the black hole.
In the sub-Eddington regime (the case of larger black hole masses), the peak
emerging luminosity is exactly the characteristic luminosity we estimated previously
(eqn.~\ref{eq:charlum}).  The black hole mass dividing the super- and sub-Eddington regimes is
\begin{equation}
M_{BH{\rm therm}} = 9.0 \times 10^7(\eta/0.1)^{2/3} (\dot m/\dot m_0)^{2/3}
                            {\cal M}_*^{(1+3\xi)/2}\kf^{-1/3}\beta^{-2} M_{\odot}.
\end{equation}
In other words, $L_{\rm therm}$ has a {\it maximum} at a black hole mass
$\simeq M_{BH{\rm therm}}$.  The thermal luminosity is $\propto M_{BH}$ at smaller masses
because photon trapping limits the emergent luminosity to $L_E$;
it is $\propto M_{BH}^{-1/2}$ at higher masses as it reflects $\dot M$ in the radiatively
efficient regime.  Much as in the case of the jet luminosity, the low-mass case will
almost always be the regime relevant to the peak of the flare, but the high-mass case
may be seen as $\dot m$ decreases.  Note that our critical mass for photon-trapping,
$M_{BH{\rm therm}}$, is several times greater than the mass estimated by
\cite{ulmer99} because we have allowed for the $k/f$ factor in the expression for $R_T$.

\begin{figure}

\includegraphics[width=0.8\textwidth]{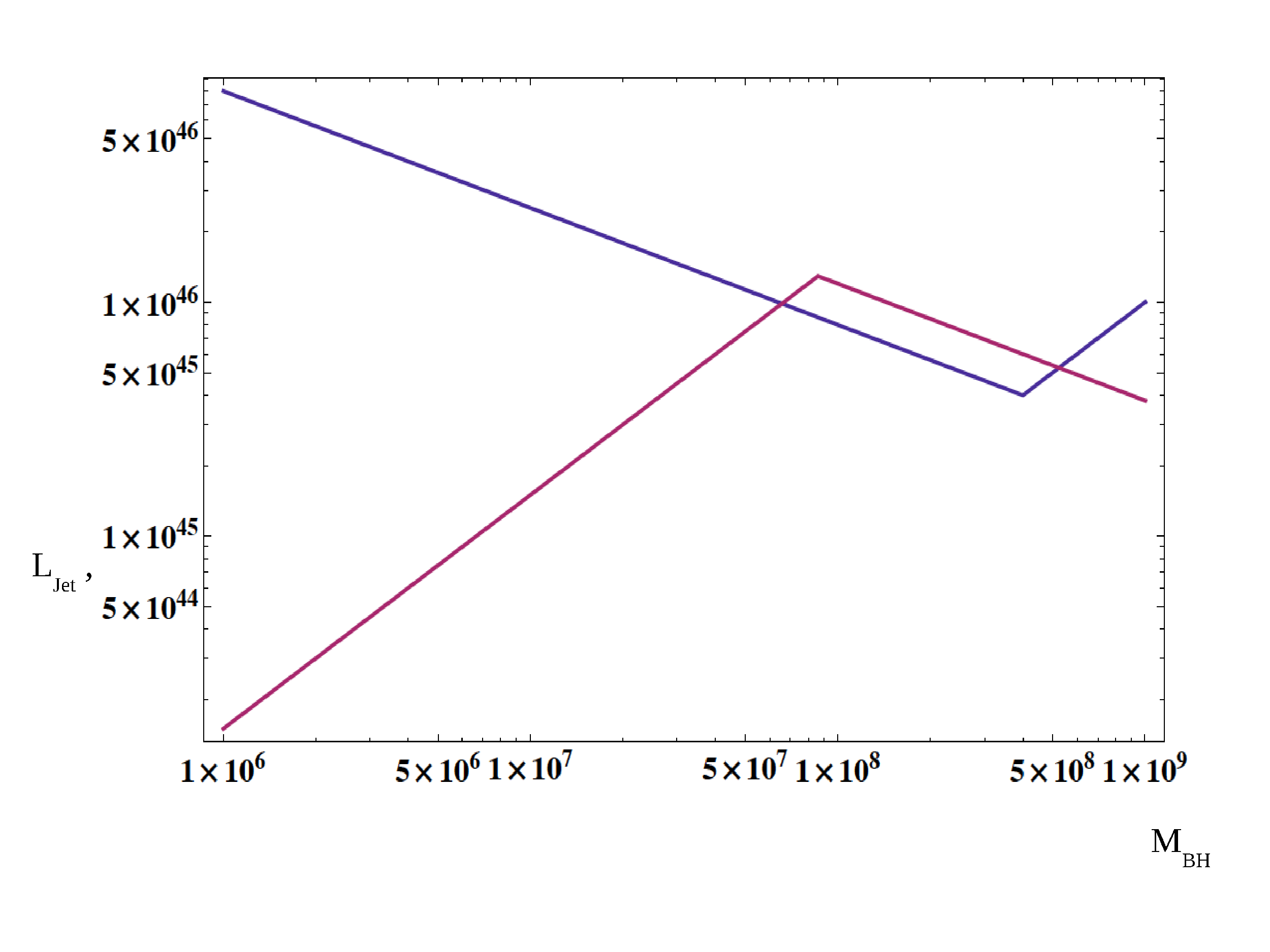}

\caption{Peak jet power (blue) and thermal (red) luminosity as a function of black hole
mass, following the prescriptions of equations~\ref{eq:ljet} and \ref{eq:ltherm}.  All
scaling factors except $M_{BH,6}$ in those expressions are set to unity.
\label{fig:L-MBH}}
\end{figure}

These trends for the peak values of $L_{\rm jet}$ and $L_{\rm therm}$ as functions of
$M_{BH}$ are illustrated in Figure~\ref{fig:L-MBH}.  The $L_{\rm jet}$ in this figure
is to some degree a maximal estimate, as it supposes both $q=1$ (i.e., little loss of accretion rate
due to outflows) and $f(a/M)=1$.  However, as the figure shows, if the black hole
rotates rapidly enough to make $f(a/M)$ not too small, for $M_{BH} \ll 10^8 M_{\odot}$,
a very large proportion of the returning mass would have to be ejected in order for
$L_{\rm jet}$ to fall to a level at which it is merely comparable with $L_{\rm therm}$.
For rapidly-spinning black holes, $L_{\rm therm} \sim L_{\rm jet}$ only when the mass is
the largest permitting any tidal disruptions at all.

Thus, the peak output is almost always dominated by the jet unless the black hole rotates
rather slowly and the black hole mass is $\sim 10^8 M_{\odot}$, a mass so large that the
tidal disruption probability is considerably less than unity.  However, whether non-thermal
radiation from the jet or thermal radiation from the disk dominates the observed spectrum
depends on the interplay of the jet radiative efficiency and beaming
with the intrinsic jet/disk ratio determined by $a/M$ and $M_{BH}$.  The jet is
favored by viewing angles along its axis, low black hole mass, high spin,
and large $\eta_{\rm jet}$; the disk is favored by viewing angles outside the favored
cone, high black hole mass, slow spin, and small $\eta_{\rm jet}$.

\subsection{Time-dependence}

Now consider what happens at later times, as matter of progressively smaller binding
energy returns to the disk.  If the mass per unit energy of tidally disrupted matter
is independent of binding energy, the accretion rate into the disk at radii
$\sim R_p$ falls $\propto (t/t_0)^{-5/3}$, where $t_0 \sim P_{\rm orb}(a_{\rm min})$
\citep{rees88,phinney89}.  \cite{lkp09} found that, depending on the degree of density
concentration in the star, the mass accretion rate might fall more shallowly than this when
$t \gtrsim t_0$, but gradually tends toward $t^{-5/3}$ when $t \gg t_0$.  If nearly all
the matter returning to the disk makes it all the way to the black hole,
the time-dependence of the accretion relevant to the disk and jet powers should match the
time-dependence of the accretion rate reaching the outer disk; if, however,
super-Eddington conditions lead to a significant fraction being expelled, the accretion
rate reaching the event horizon falls more slowly because the expelled fraction can also
be expected to diminish.

When $\dot m$ first begins to decrease, the thermal radiation hardly changes
because photon-trapping continues to limit $L_{\rm therm}$ to $L_E$.  Consequently,
in the period near and shortly after the peak luminosity, the thermal light curve
should be almost flat and only slowly roll over toward $(t/t_0)^{-5/3}$.

On the other hand, $L_{\rm jet}$ in those sources is $\propto q\dot m$, so the jet
power falls, provided the accretion rate reaching the black hole is truly
super-Eddington.  The rate of decrease may be slower than $t^{-5/3}$ if $q$ increases
as $\dot m$ falls.  With declining $\dot m$, the black hole mass at which the flow
switches from non-radiative to radiative also decreases.  Consequently,
even though virtually every source begins in a non-radiative state, those in
which the black hole mass is relatively large may ultimately become efficient
radiators at later times.  When that changeover occurs, the jet power remains
constant, while the thermal luminosity begins to fall.  This timescale, when the
accretion rate and therefore the thermal luminosity fall below Eddington, we call
$t_{\rm Edd}$.  It should be emphasized
that all of these remarks pertain to the {\it expectation value} for the jet
luminosity; relativistic jets are generically unsteady, so fluctuations at the
order-unity level should be expected around all these trends.

\begin{figure}

\includegraphics[width=0.8\textwidth,angle=90]{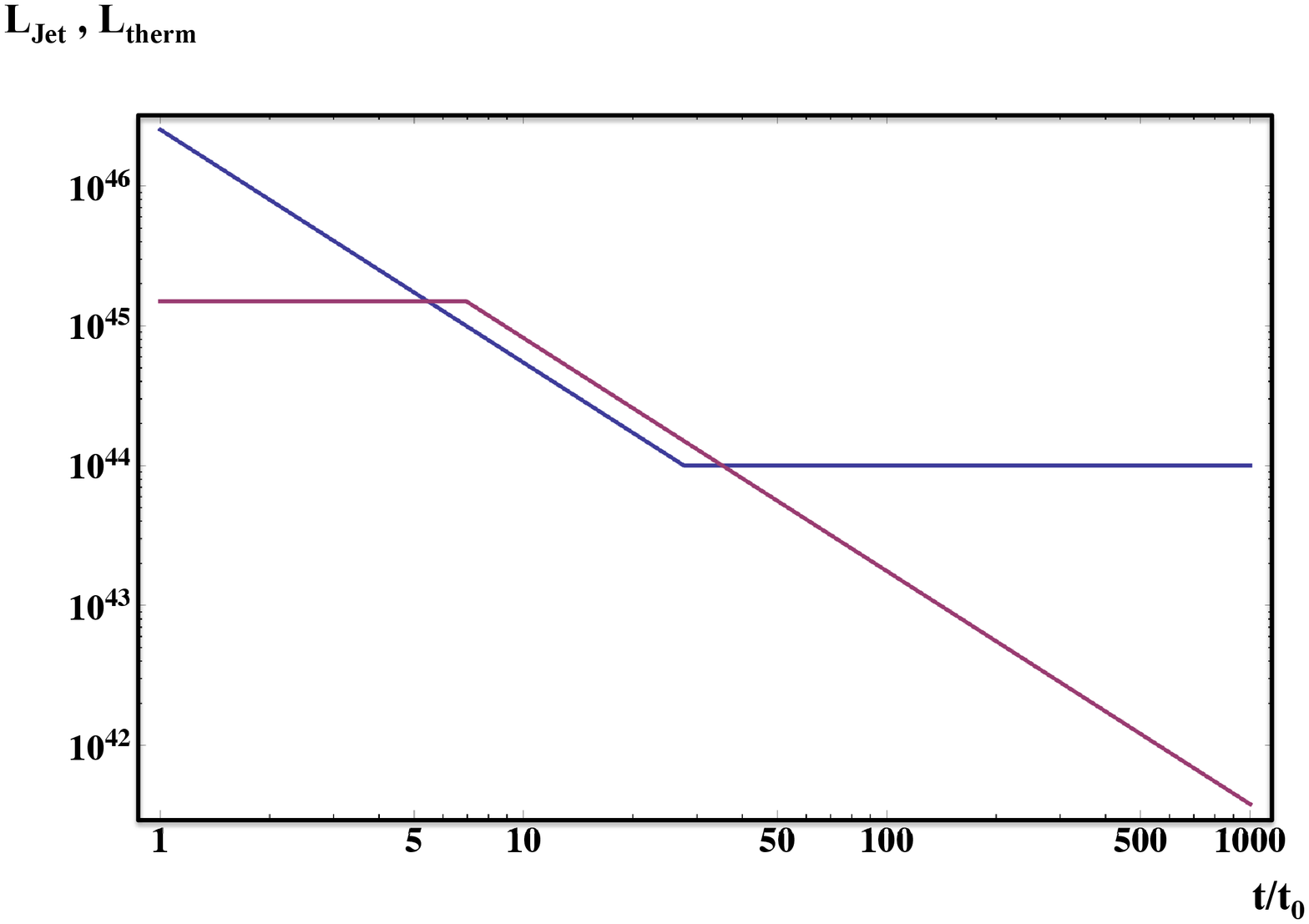}

\caption{Jet power (blue) and thermal (red) luminosity as functions of time for $M_{BH}
= 1 \times 10^7 M_{\odot}$.  The Eddington timescale $t_{\rm Edd}$ for these parameters
is $7 t_0$.  As in Figure~\ref{fig:L-MBH}, all scaling factors except
$M_{BH,6}$ and $\dot m/\dot m_0$ are set to unity. 
\label{fig:L-t}}
\end{figure}

Figure~\ref{fig:L-t} gives a schematic view, beginning at the time of peak accretion
rate, of what might be expected in terms of the light curves for the jet power (before
allowance for beaming and radiative efficiency) and the thermal luminosity.  For the
parameter values chosen ($M_{BH,6} = 10$, all other scaling parameters unity),
$L_{\rm jet}$ falls to the level of $L_{\rm therm}$ at almost the same time, $t \simeq 7 t_0$,
as $L_{\rm therm}$ enters the sub-Eddington regime and also begins to decline.  From that
time to $t \simeq 30 t_0$, both fall together, maintaining similar power levels.  Finally,
after $t \simeq 30t_0$ (i.e., a time larger by $\eta^{-3/5} \simeq 4$ than the time at which the
thermal luminosity begins to decline), the jet luminosity stabilizes, while
$L_{\rm therm}$ continues to fall.

The ratio $L_{\rm jet}/L_{\rm therm}$ as a function of time for fixed black hole
mass can be described much more simply.  In rough terms, it is just
\begin{equation}
\frac{L_{\rm jet}}{L_{\rm therm}} \propto
          \frac{\max(1,q\dot m) f(a/M)}{\min[1,\eta(a/M)\dot m]},
\end{equation}
where $\eta(a/M)$ is the intrinsic (i.e., without photon trapping)
radiative efficiency for a given spin parameter.  At high accretion rate, the ratio
is $\propto q\dot m$; in the transition range where photon trapping is marginal, the
ratio may change relatively slowly; at low accretion rate, the ratio is
$\propto \dot m^{-1}$.  In any particular event, $\dot m$ rises rapidly (on a
timescale $\sim t_0$) at first, but then declines slowly.  The ratio of {\it observed}
jet luminosity to thermal luminosity follows the same trend, modulated by any
evolution of $q{\cal B}\eta_{\rm jet}$.

Eventually, the accretion rate falls so low that the disk is no longer radiation-dominated,
and our estimate for the pressure in the inner disk is no longer valid.
Assuming that this occurs when the gas pressure becomes comparable to the radiation pressure
at $r \simeq 10r_g$, the critical accretion rate is
$\dot m \simeq 0.05 (\eta/0.1)^{-1}(\alpha/0.1)^{-1/8} M_{BH,6}^{-1/8}$.  If the accretion
rate declines as $(t/t_0)^{-5/3}$, it happens at a time
\begin{equation}
t_{\rm g} \simeq 1300 t_0 \beta^{-1/5} (\eta/0.1)^{3/5} (\alpha/0.1)^{3/40}
                 M_{BH,6}^{-33/40} {\cal M}_*^{3(1+3\xi)/10}.
\end{equation}
After this time, although the thermal luminosity continues to decline in proportion to the
accretion rate, the midplane pressure (and therefore jet luminosity) declines
$\propto \dot m^{4/5}$.  This break time is quite late in the development of the flare for
low-mass events, but when the black hole mass is relatively large, it could take place
as early as $\simeq 30 t_0$.

%   D. Note that short-lived tidal disruption jets are mostly hard photon emitters because the
%      jet working surface is still so close to the black hole that it's strongly self-absorbed
%      at radio synchrotron frequencies.

\section{Constraining tidal disruption parameters: the case of J2058.4+0516}
\label{sec:j2058}

Although the theory of jet and disk emission from tidal disruptions as we have presented
it contains a large number of free parameters, in any given event there are also potentially
a sizable number of observable properties that may be used to constrain these parameters.
In nearly every event, it is possible to measure the characteristic time $t_0$.
Optical and/or ultraviolet observations often give an indicator of the peak thermal disk
luminosity, $L_{d0}$; we describe this as only an ``indicator" because, as we discuss
below, there is likely to be a sizable---and uncertain---bolometric correction.
When there is hard X-ray emission, the peak jet luminosity $L_{j0}$ can also be obtained;
if an event can be followed long enough, it may also be possible to measure two times
related to the transition from super- to sub-Eddington accretion: the time $t_{\rm jet}$
at which the jet luminosity flattens, and the time $t_{\rm disk}$ at which the disk luminosity
begins to diminish.  In this section, we will lay out a general formalism for
using these observables as parameter constraints and then apply that method to a specific
example, Swift J2058.4+0516.

As shown by \cite{lkp09}, the accretion rate peaks at a time $P_{\rm orb}(a_{\rm min})$
past pericenter passage (see eqn.~\ref{eq:porb_min}), and diminishes thereafter as a power-law
in time.  If we identify the time of peak flare luminosity $t_0$ with that orbital period, we
obtain the constraint
\begin{equation}
t_{0,d} \simeq 5.8 {\cal M}_*^{(1-3\xi)/2} M_{BH,6}^{1/2} \kf^{1/2} \beta^3,
\end{equation}
where $t_0$ is measured in days.

Because the peak accretion rate is likely to be super-Eddington for almost the entire
range of possible black hole masses, we expect the peak disk luminosity to be close to
the Eddington luminosity.  Unfortunately, however, the characteristic temperature
of thermal disk radiation at the Eddington luminosity is $\sim 1 \times 10^6 M_{BH,6}^{-1/4}$~K,
indicating that the bulk of the light may emerge in the EUV, where direct measurements
are very difficult.  Consequently, a bolometric correction that could well be
$\sim O(10)$ or greater must be applied to any observable measure of the disk luminosity.
Unfortunately, given our current limited understanding of disk spectra even when the accretion
rate is sub-Eddington, not to mention potential dust extinction in the host galaxy or possible
reprocessing in a wind \citep{strubbe11}, a sizable uncertainty must be attached to any such
correction.  For this reason, we give higher priority to the use of other observables.

In our description of the time-dependence of the jet and disk luminosities,
the transition from super- to sub-Eddington behavior occurs earlier for the disk than for
the jet.  The ratio of the accretion rates at these two timescales is $\sim \eta$, the
radiative efficiency of the disk in the trans-Eddington regime.   Consequently, the ratio
of these two timescales primarily constrains the black hole spin, but not any of the other
parameters.  For the other parameters, no additional information is gained by using both
timescales; either one will do.

Suppose, then, that we choose to use the timescale at which the jet luminosity flattens
as a function of time, $t_{\rm jet}$.  This timescale is typically a few times $t_{\rm Edd}$.
For generality, let the accretion rate scale with time
as $(t/t_0)^{-n}$, where $n \simeq 5/3$ is expected.  Then we find that
\begin{equation}
t_{\rm jet,d} \simeq 5.8 (8\times 10^3)^{1/n} {\cal M}_*^{[1 + 1/n + 3\xi(1/n-1)]/2} M_{BH,6}^{(1-3/n)/2}
              \kf^{1/2}\beta^{3(1 -1/n)},
\end{equation}
where $t_{\rm jet}$ is likewise scaled in days.
Combining this timescale constraint with the one based on the characteristic flare timescale,
we may solve for the black hole mass and the penetration factor:
\begin{equation}
M_{BH,6} \simeq 4.6 \times 10^4 {\cal M}_* t_{0,d}^{n-1} t_{\rm jet,d}^{-n}
\end{equation}
and
\begin{equation}
\beta \simeq 0.093 {\cal M}_*^{\xi/2 - 1/3} \kf^{-1/6} t_{0,d}^{1/2-n/6} t_{\rm jet,d}^{n/6} .
\end{equation}

Lastly, using both of these estimates, the (numerous) parameters governing the jet luminosity
may be constrained:
\begin{equation}\label{eq:qbeta}
{\cal B}\eta_{\rm jet} q f(a/M)/(\beta_h\alpha) \simeq 0.022 L_{jet,46} {\cal M}_*^{-1} t_{0,d},
\end{equation}
where $L_{jet,46}$ is the peak luminosity of the jet in units of $10^{46}$~erg~s$^{-1}$.

\begin{figure}[h]
\includegraphics[width=\textwidth]{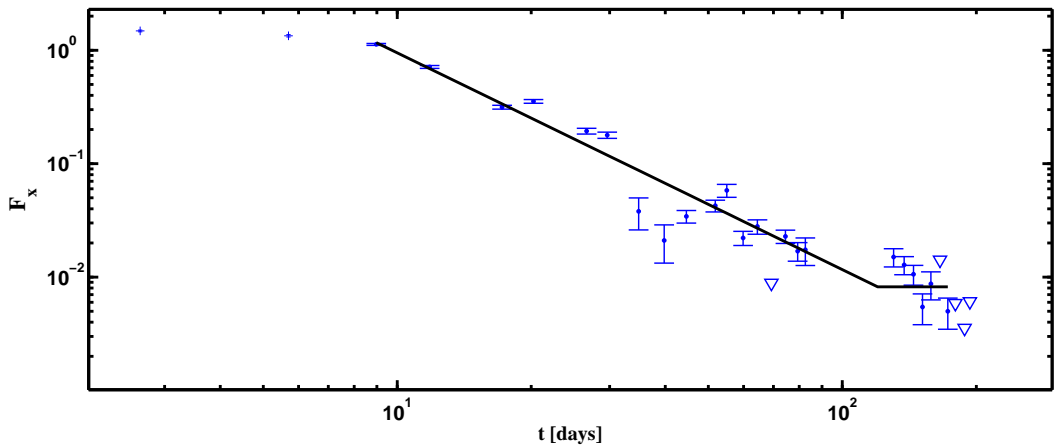}

\caption{Long-term Swift XRT light curve in the 0.5--10 keV band for J2058
as of 3 January 2012 (last data point 7 December 2011), combining WT and PC data
(data drawn from http://www.swift.psu.edu/monitoring).  $F_x$ is Swift counts per
second.  Triangles show $1\sigma$ upper bounds.  The solid line is the result of
a $\chi^2$-squared fit to the lightcurve for $t > 7$~d.
\label{fig:lc}}
\end{figure}

The recently-discovered flare source J2058.4+0516 \citep{cenko11} may be an example
of exactly the sort of tidal disruption event to which this formalism is applicable.
Its lightcurve is shown in Figure~\ref{fig:lc}.  For $\simeq 10$~d, its flux stayed
nearly constant; for the next three months, the flux declined roughly as a power-law
in time.  Around $t \sim 100$~d, the decline became shallower; unfortunately, the
flux at that point was already near Swift's detection limit, so the lightcurve beyond
$\simeq 120$~d is a mix of detections and upper bounds whose $1\sigma$ limits are not
very different from the level of some of the last detections.  To describe this
lightcurve parametrically, we fit the data for $t > 7$~d to a model of the form
$F(t) = A t^{-n}$ for $t < t_b$ and $F(t) = B$ for $t \geq t_b$.  The minimum in
$\chi$-squared was found for $n = 1.9$ and $t_b \simeq 120$~d; this best-fit model
is shown as the solid line in Figure~\ref{fig:lc}.  That $\chi^2$ is reduced by
a value of $t_b$ within the span of the data supports our prediction that the jet
luminosity should become nearly constant after an initial period of decline.

To use the procedure just outlined, we take $t_0 = 10$~d, as that appears to be the
point at which power-law decline begins.  Following the results of our fit to the lightcurve,
we set $t_{\rm jet} = 120$~d, uncertain as that identification may be, and $n=1.9$ rather 
than 5/3.  The difference in the parameter inferences due to the latter choice
are probably smaller than the intrinsic error in the method.  Using these numbers, we find
\begin{equation}
M_{BH,6} \simeq 40 {\cal M}_*
\end{equation}
and
\begin{equation}
\beta \simeq 0.6 {\cal M}_*^{\xi/2 - 1/3} \kf^{-1/6}.
\end{equation}
That is, we estimate that the black hole is in the upper range of masses at which tidal disruptions
of main sequence stars can occur, and the penetration factor is quite modest.  The expected
mass might be a bit smaller if the stellar mass is less than solar; in that case, $\xi \simeq 0.2$,
and $\beta$ would increase.

The luminosity in this flare appears to be dominated by hard X-rays (photon energies
$\gtrsim 10$~keV) and peaked (in nominal isotropic terms) at
$\sim 3 \times 10^{47}$~erg~s$^{-1}$ (or perhaps a factor of a few more when contributions
from still harder X-rays are considered).  The constraint on the jet parameters
(eqn.~\ref{eq:qbeta}) then becomes
\begin{equation}
q{\cal B}\eta_{\rm jet}f(a/M)/\left(\beta_h\alpha\right) \simeq 
             6 {\cal M}_*^{-1} .
\end{equation}
For self-consistency, the jet must be reasonably beamed and efficient, and the black
hole must be spinning rapidly enough to make $f(a/M)$ not too small.  If ${\cal M}_* < 1$,
the beaming, jet efficiency, spin, etc. must be somewhat greater.

Although the available optical/UV observations are only very rough guides for further inference,
they are consistent with these values. Optical photometry suggests a
steeply rising spectrum in the observed frame between the $g^\prime$-band and the
$u$-band \citep{cenko11}.  The $u$-band flux on its own translates to a luminosity
$\nu L_\nu \simeq 1.1 \times 10^{45}$~erg~s$^{-1}$ given the object's redshift
($z=1.18$) and assuming a flat cosmology with $H_0 = 70$~km~s$^{-1}$~Mpc$^{-1}$.
If $L_{\rm therm}$ is Eddington-limited and isotropic, this luminosity alone
requires a black hole mass $\gtrsim 7 \times 10^6 M_{\odot}$.  However, the
mass estimated in this fashion is proportional to the bolometric correction,
and, as we have already noted, this may well be at least an order of magnitude.
If so, the mass required would be in the range just estimated on the basis of
$t_0$ and $t_{\rm jet}$.

Lastly, we consider what information may be gleaned from the time-dependence of the optical/UV
flux.  \cite{cenko11} report $g^\prime$-band optical photometry taken within 10~d of the
flare's initiation and $\simeq 10$~d, $\simeq 30$~d, and $\simeq 40$~d later.  They also
report $u$-band magnitudes at $\simeq 10$~d and $\simeq 40$~d.  In $g^\prime$, there was
no detectable dimming until $\simeq 30$~d after the flare, but at $\simeq 40$~d, the latest
time reported, the flux had diminished by $\simeq 0.5$~mag, $\simeq 40\%$.  The $u$-band
flux drops about 0.5~mag from $\simeq 10$~d to $\simeq 40$~d.  Thus, in rough terms, the
optical/UV luminosity appears to have varied very little over the first month or so, even
though the X-ray luminosity fell by a factor $\sim 8$; such behavior is in good agreement
with our prediction that the disk output should remain steady while the jet power falls
during the super-Eddington phase.  If the drop between 30~d and 40~d
marks the disk transition to sub-Eddington behavior, the factor $\sim 3$--4 between that
timescale and the possible timescale of the jet transition is also consistent with our suggestion
that the accretion rate at the time of disk transition is a factor $\sim \eta^{-1}$ greater
than at the time of jet transition.

\section{Conclusions}
\label{sec:conc}
We have demonstrated that tidal disruption of main sequence stars by rotating
black holes can lead to the launching of a powerful relativistic jet.  Because
the peak accretion rate in most of these events is likely to be
super-Eddington, a regime in which photon trapping is significant, the thermal
luminosity may be held to no more than roughly the Eddington luminosity; that is,
the thermal luminosity is suppressed well below what the accretion rate would
predict on the basis of the usual relativistic radiation efficiency.  As a result,
the thermal lightcurve (in bolometric terms) should have a peak that is rather
flatter than the curve describing the accretion rate.  On the
other hand, the jet power suffers no such suppression, so it can exceed the thermal
luminosity of the accretion flow.

As the accretion rate declines, the jet power diminishes
until the sub-Eddington regime is reached; after that point, we expect that its power
changes little until the accretion rate is so low that the disk is no longer
radiation-dominated.   By contrast, the thermal luminosity can be expected
to remain roughly constant at roughly the Eddington luminosity as the accretion
rate falls, diminishing with the classic $t^{-5/3}$ scaling only after the flow
becomes sub-Eddington and radiatively efficient.  The time at which the thermal
luminosity begins to fall we have named $t_{\rm Edd}$; the time at which the
jet lightcurve levels out is a few times $t_{\rm Edd}$.

It follows that in many main sequence tidal disruptions, the jet luminosity may be
comparable to or greater than the thermal luminosity.  Particularly when the
black hole rotates rapidly, the non-thermal jet photon luminosity may substantially exceed
the thermal disk luminosity even after allowing for comparatively inefficient radiation
of the jet's kinetic power.  The degree to which we see this non-thermal luminosity
depends, of course, on our viewing angle.  When it is favorable, the non-thermal
emission might be significantly larger than the thermal.

Particularly when the reach of the survey extends only to low redshifts, beaming also
enhances the accessible population.  Although the fraction of sources we see is reduced by
${\cal B}$, the luminosity distance to which we can see them (for negligible k-correction)
is increased by ${\cal B}^{1/2}$, implying an increase in accessible comoving volume of
${\cal B}^{3/2}/(1+z_{\rm max})^3$, for $z_{\rm max}$ the greatest redshift at which the
luminosity of interest can be detected (provided $z_{\rm max} \lesssim 1$, so that
cosmological corrections are small).  Thus, if ${\cal B} \sim 100$ (as from relativistic
beaming with a Lorentz factor $\Gamma \sim 10$), beaming results in an increase in the
number of detectable sources by a factor $\sim O(10)$ provided the population has
redshifts smaller than $\simeq 1$.
%More precisely, if neither the stellar mass function nor,
%more importantly, the black hole mass function evolves over the range of relevant
%redshifts, the number of sources over the whole sky with flux greater than $F$ is
%\begin{equation}
%N(>F) = \frac{4\pi}{3} \int \, dM_* \, \frac{dN}{dM_*}\int \, dM_{BH} \, \frac{dN}{dM_{BH}}
%  {\cal B}^{-1} \left[{\cal B}L(M_{BH},M_*)/(4\pi F)\right]^{3/2}
%            \left(1 + z_{\rm max}\right)^{-3}.
%\end{equation}
%The maximum redshift is, of course, also a function of $M_{BH}$ and $M_*$ through its
%dependence on ${\cal B}L(M_{BH},M_*)$.

For these reasons, it may therefore be desirable to augment traditional UV-based methods
of searching for stellar tidal disruptions (e.g., \cite{gezari08}) with other methods more
sensitive to jet radiation, perhaps based on the Swift system of all-sky monitoring in hard
X-rays coupled to 2--10~keV X-ray follow-up (e.g., using possible future missions such
as the Space-based multi-band astronomical Variable Objects Monitor [SVOM: \cite{svom11}]
or the Advanced X-ray Timing Array [AXTAR: \cite{axtar10}]).  Because tidal disruption
flares last much longer than the $\gamma$-ray bursts for which Swift was designed, the
follow-up (for events involving main sequence stars) can be delayed by as much as $\sim 1$--10~d.

Applying this analysis to the recently-discovered example of Swift J2058, we find that
this event is consistent with a tidal disruption of a main sequence star of roughly solar
mass by a black hole of mass $\sim 4 \times 10^7 M_{\odot}$.  Its hard spectrum suggests
a jet of the sort we have described, while its characteristic timescale and X-ray
luminosity are likewise consistent with this picture.  Our model predicts that the jet
luminosity should fall when the accretion rate begins to fall, but become constant later
in the flare when the accretion rate becomes sub-Eddington; the observed lightcurve
suggests this sort of behavior.  Published optical photometry is also consistent with
our prediction that the optical/UV thermal luminosity should initially be almost independent
of time, although this consistency is weakened by the large, and plausibly time-dependent,
bolometric correction that must be applied.
Observations using more sensitive detectors might be able to test these predictions
more definitively.

Finally, we remark that, although the timescales of Galactic black hole transients
are such that we can observe a rich phenomenology of spectral states related to
the magnitude of the accretion rate, the corresponding timescales for supermassive
black holes in AGN are far too long to permit human observation.
Main sequence tidal disruptions, on the other hand, present us with a remarkable
laboratory for studying how the properties of accretion onto supermassive black holes
depend on accretion rate.  Beginning well above Eddington, the accretion rate in these
systems declines on timescales from weeks to years, and in a fashion we (at least
partially) understand.  We can hope that further study of these events may shed
light on the relation between the magnitude of the accretion rate and thermal disk
emission, jets, and coronal X-rays.

\acknowledgements{This work was partially supported by NSF grants AST-0507455 and
AST-0908336 (JHK) and by an ERC advanced research grant and the ISF center for High
Energy Astrophsics (TP).  We thank Michael Stroh for help with the Swift data.}

%%%%%%%%%%%%%%%%%%%%%%%%%%%%%%%%%%%%%%%%%%%%%%%%%%%%%%%%%%%%%%%%%%%%%%
% BIB
%%%%%%%%%%%%%%%%%%%%%%%%%%%%%%%%%%%%%%%%%%%%%%%%%%%%%%%%%%%%%%%%%%%%%%
%\bibliography{bib}

%%%%%%%%%%%%%%%%%%%%%%%%%%%%%%%%%%%%%%%%%%%%%%%%%%%%%%%%%%%%%%%%%%%%%%
%%%%%%%%%%%%%%%%%%%%%%%%%%%%%%%%%%%%%%%%%%%%%%%%%%%%%%%%%%%%%%%%%%%%%%

\end{document}